\documentclass{cernrep}
\usepackage{texnames}
\usepackage[T1]{fontenc}
\newcommand{\emiss}{\mbox{$\not \!\!E$}}
\newcommand{\pmiss}{\mbox{$\not \! p$}}
\newcommand{\ptmiss}{\mbox{$\vec {\not \! {p_T}}$}}
\newcommand{\ppmiss}{\mbox{$\vec {\not \! p_{_\parallel}}$}}
\newcommand{\ppb}{\mbox{$\vec p_{{_\parallel} b}$}}
\newcommand{\ppar}{\mbox{$\vec{p_{_\parallel}}$}}
\newcommand{\pt}{\mbox{$\vec{p_T}$}}

\begin{document}
\title{A Method for Deriving Transverse Masses Using Lagrange Multipliers}

\author{Eilam Gross and Ofer Vitells}

\institute{Weizmann Institute, Rehovot, Israel \\
eilam.gross@weizmann.ac.il,ofer.vitells@weizmann.ac.il}

\maketitle 

\begin{abstract}
We use Lagrange multipliers to extend the traditional definition of
Transverse Mass used in experimental high energy physics. We
demonstrate the method by implementing it to derive a new Transverse
Mass that can be used as a discriminator to distinguish between top
decays via a charged W or a charged Higgs Boson.

\end{abstract}

\section{Introduction}

If a particle decay products are not fully measurable  due to the
presence of Neutrinos in the final states or energy flowing in the
direction of the beam pipe, it is impossible to fully reconstruct
the mass of the decaying particle. In proton-proton collisions, the
missing momentum in the $z$ direction is a complete unknown and
cannot be deduced by energy-momentum balance considerations (unlike
the missing transverse momentum). However, using known constraints
on unmeasurable physical quantities (usually from energy-momentum
conservation) it is possible to find an upper or lower bound on the
decaying particle mass. In section \ref{sec:W} we derive the
classical $W$ transverse mass  when the $W$ Boson decays to a Lepton
and a Neutrino, $W\rightarrow \ell\nu$ \cite{bib:wmt1}\cite{bib:wmt2}. The $W$ mass bounds this
Transverse mass from above. In section \ref{sec:WH} we derive a
different expression for the $W$ "Transverse Mass"  in top decays
involving multi-neutrinos final states, $t\rightarrow Wb\rightarrow
\tau\nu b\rightarrow\ell\nu\bar\nu \nu b$. Here the mass is bounded
from below. We then implement the same expression to discriminate a
top decaying via a $W$ and a charged Higgs Boson. Charged Higgs
Bosons are smoking guns for the existence of theories beyond the
Standard Model, like the Minimal Supersymmetric extensions of the
Standard Model (MSSM) which contain two charged Higgs Bosons
\cite{bib:MSSM}. In section \ref{sec:WWHH} we demonstrate how to
implement the method to derive additional kinematical bounds. We
then conclude.

\section{Notation}
\label{notation}
 For convenience we adopt the following notation:

$p$ is a
  4-vector
\begin{equation}
  p\equiv(\ppar ,\pt)
\end{equation}
 with
\begin{equation}
  \pt=(p_x,p_y)
\end{equation}
 and
 \begin{equation}
 \ppar=(E,p_z)
 \end{equation}
  satisfying
\begin{eqnarray}
 \label{eq1}
  p^2=m^2=\ppar^2-\pt^2 \\
  \ppar^2=E^2-p_z^2 \\
  \pt^2=p_x^2+p_y^2.
\end{eqnarray}
The missing momentum is denoted by $\pmiss$.

\section{The case of one Neutrino: deriving the classical $W$ transverse mass}
\label{sec:W}
 Suppose the outcome of a proton-proton collision is a
$W$ Boson decaying into a lepton and a Neutrino, $W\rightarrow
\ell\nu$.
 The mass of the $W$ is given by $M^2=( p_{\ell}+\pmiss)^2$.
 Assuming one can measure the transverse missing momentum,  we
 consider $\pmiss^Z$ and $\emiss$
  as two  unknown quantities satisfying the constraint
 $\pmiss^2=0$. Using Eq. \ref{eq1}
 the mass can be expressed as
 \begin{equation}
 \label{eqM}
 M^2=(\ppar_\ell+\ppmiss)^2-(\vec p_{T\ell}+\ptmiss)^2
 \end{equation} and the constraint as
 \begin{equation}
 \label{eqC}
 \ppmiss^2-\ptmiss^2=0.
 \end{equation}
 To find an extremum to the squared mass (\ref{eqM}) under the
 constraint (\ref{eqC}) we use the method of Lagrange multipliers.
 Since the transverse momenta are orthogonal to the parallel momenta
 we find
\begin{equation}
\frac{\partial}{\partial \ppmiss} ( ( \ppar_\ell+\ppmiss )^2-\lambda(\ppmiss )^2 )\mid_{\lambda=\lambda_0,\ppmiss=\ppmiss_0}=0
\end{equation}
We find
\begin{equation}
(\ppar_\ell+\ppmiss_0)=\lambda_0\ppmiss_0 . \label{eqB}
\end{equation}
 Using (\ref{eqC}) the solution $\lambda_0$ for $\lambda$ is
\begin{equation}
(\lambda_0-1)^2=\frac{{\ppar_\ell}^2}{\ptmiss^2}
\end{equation}
Approximating the lepton to be massless $ ( \ppar_\ell^2=\vec
p_{T\ell}^2)$ , we find
\begin{equation}
\lambda_0=1+\frac{p_{T\ell}}{\not \! p_T} \label{eqL}
\end{equation}
where $p_{T\ell}$ and $\not \! p_T $ are the magnitudes of
the transverse lepton and missing momenta. Plugging $\lambda_0$ into
(\ref{eqB}), and using (\ref{eqC}) we find the extremum for the
squared mass (Eq. \ref{eqM}) denoted by $M_T$. It is given by
\begin{equation} M_T^2=(\not \! p_T+p_{T\ell})^2-(\vec
p_{T\ell}+\ptmiss)^2 \label{eqMM1}
\end{equation}
which can also be expressed as
\begin{equation}
M_T^2=2\not \! p_T p_{T\ell}(1-cos\psi) \label{eqMM2}
\end{equation}
where $\psi$ is the angle between the lepton and the missing
momentum in the transverse plane. Equations \ref{eqMM1} and
\ref{eqMM2} are the familiar forms of the $W$ transverse mass \cite{bib:wmt1}\cite{bib:wmt2}.

\begin{figure}[h!]
\begin{center}
\includegraphics[width=0.5
\textwidth]{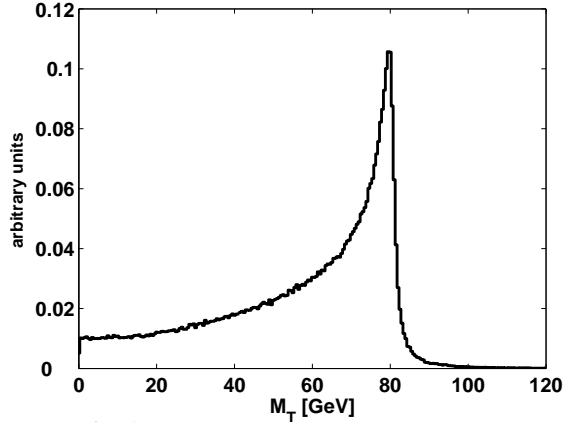}
\vspace*{-0.5cm}
\caption{The $W$ transverse
mass distribution}
\label{fig:FigA}
\vspace*{-0.7cm}
\end{center}
\end{figure}

Figure \ref{fig:FigA} shows the distribution of the $W$ transverse
mass, using partons generated with Pythia \cite{bib:pythia}. One
can clearly see that this distribution has a sharp upper bound at
$M_T=M_W$, as expected.

\section{The case of multi-neutrinos: deriving a Chraged Higgs transverse mass}
\label{sec:WH}
 Here we consider a final state with a lepton and
three Neutrinos. Such a final state can occur in top decays. For
example
 $t\rightarrow
Wb\rightarrow \tau\nu b\rightarrow\ell\nu\bar\nu \nu b$. One
interesting case is when the $W$ Boson is replaced by the charged
Higgs Boson $H^+$, which occurs in Mimnimal Supersymmetric
extensions of the Standard Model (MSSM). Here one would like to tell
$W$ mediated decays from Higgs mediated decays. One possible
discriminating variables is the transverse mediating particle mass
which is derived in this section.

The difference between this case and the previous one (section
\ref{sec:W}) is that here the missing energy-momentum is a result of
a few Neutrinos and one cannot use the constraint (Eq. \ref{eqC}).
Here $ \pmiss^2=\ppmiss^2-\ptmiss^2\neq 0$. However, this constraint
can be replaced by a constraint given by the parent particle mass,
i.e. the top quark mass
\begin{equation}
M^2_{top}=( p_\ell+\pmiss+ p_b)^2 \label{eqT}
\end{equation}
where $p_b$ is the bottom-quark 4-momentun.

 The mediating particle
mass which is subject to minimization under the constraint
(\ref{eqT}) is given by
\begin{equation}
M^2=( p_\ell+\pmiss)^2.
\label{eqU}
\end{equation}
Repeating the same procedure as in section \ref{sec:W} we find
\begin{equation}
(\ppar_\ell+\ppmiss_0)=\lambda_0(\ppar_\ell+\ppmiss_0+\ppb).
 \label{eqV}
\end{equation}
Where $\lambda_0$ and $\ppmiss_0$ are minimizing Eq. \ref{eqU}. With
the help of the constraint (\ref{eqT})
 we find the extremum denoted by $M_T$,
\begin{equation}
M_T^2=\left (\sqrt{ M_t^2+(\vec p_{T\ell}+\vec
p_{Tb}+\ptmiss)^2}-p_{Tb} \right )^2-\left ( \ptmiss+\vec p_{T\ell}
\right)^2 \label{eqX}
\end{equation}
where $p_{Tb}$ is the scalar magnitude of the bottom quark
transverse momentum.

\begin{figure}[h!]
\begin{center}
\includegraphics[width=0.5
\textwidth]{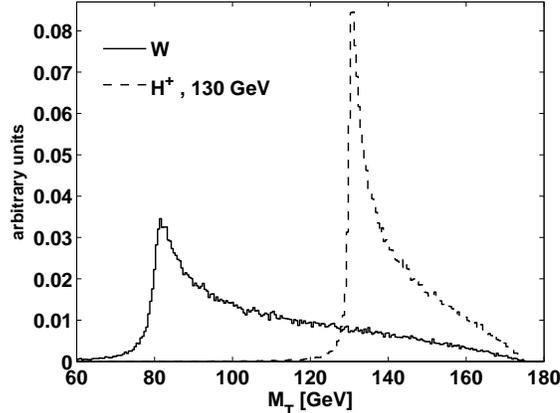}
\vspace*{-0.5cm}
\caption{The new transvesre mass distribution for a mediating $W$ Boson
(full line) and a 130 $GeV$ Charged Higgs Boson (dashed line)}
\label{fig:med}
\vspace*{-0.7cm}
\end{center}
\end{figure}

Figure \ref{fig:med} shows the parton distribution of this
transverse mass (Eq. \ref{eqX}) for two cases. A mediating $W$ Boson
(full line) and a mediating Charged Higgs Boson with a hypothetical
mass of 130 $GeV/C^2$. One can clearly see the power of this method
to distinguish between the two possible top-decays. Note also that
in this case the distributions have a threshold and a Jacobian peak at the mediating particle mass (the $W$ or the charged Higgs mass).

\section{The general case: other applications}
\label{sec:WWHH} In the general case let $F(\ppmiss,\ptmiss,....)$
be a kinematical variable which is a function of the unknown
$\ppmiss$ and other known variables. Let $f(\ppmiss,\ptmiss,....)=0$
be a constrained on the variables involved.
 Define
\begin{equation}
G(\lambda,\ppmiss,\ptmiss,....)=F(\ppmiss,\ptmiss,....)+\lambda
f(\ppmiss,\ptmiss,....)
\end{equation}
Solve for $\lambda$ and $\ppmiss$ that minimizes or maximizes
$G(\lambda,\ppmiss,\ptmiss,....)$, i.e.
\begin{equation}
\frac{\partial}{\partial \ppmiss}
G(\lambda,\ppmiss,\ptmiss,....)|_{\lambda=\lambda_0,\ppmiss=\ppmiss_0}=0
\end{equation}
The solution gives a relation
\begin{equation}
g(\lambda_0,\ppmiss_0,\ptmiss,....)=0
\end{equation}
 The resulting "transverse" $F$ is then
\begin{equation}
F_T=F(\ppmiss_0,\ptmiss,....)
\end{equation}


\section{Conclusion}
We have shown how to construct discriminators useful for Hadron
colliders where only the transverse missing energy can be measured.
These discriminators are an extension of the traditional W
transverse mass used in High Energy Physics and we derive them using
Lagrange multipliers.

\section{Acknowledgements}

One of us (E.G.) is obliged to the Benoziyo center for High Energy
Physics for their support of this work. This work was also supported
by the Israeli Science Foundation(ISF), by the Minerva Gesellschaft
and by the Federal Ministry of Education, Science, Research and
Technology (BMBF) within the framework of the German-Israeli Project
Cooperation in Future-Oriented Topics(DIP).

\end{document}